\documentclass[aps,prl,twocolumn,superscriptaddress]{revtex4-1}

\usepackage{todonotes} 
\usepackage{graphicx}
\usepackage{CJK}
\usepackage{dcolumn}
\usepackage{bm}
\usepackage[colorlinks=true,bookmarks=false,citecolor=blue,linkcolor=blue,hyperfootnotes=true,urlcolor=blue]{hyperref}
\usepackage{todonotes}
\usepackage{color}
\usepackage{comment}
\usepackage{sidecap}
\usepackage{nicefrac}
\def\mathbi#1{\ensuremath{\textbf{\em #1}}}

\begin{document}

\title{Spontaneous Magnetic Superdomain Wall Fluctuations in an Artificial Antiferromagnet}

\author{X. M. Chen}
\email{xmchen@lbl.gov}
\affiliation{Advanced Light Source, Lawrence Berkeley National Laboratory, Berkeley, CA 94720, USA}
\affiliation{Department of Electrical and Computer Engineering, University of Kentucky, Lexington, KY 40506, USA}

\author{B. Farmer}
\author{J. S. Woods}
\affiliation{Department of Physics, University of Kentucky, Lexington, KY 40506, USA}

\author{S. Dhuey}
\affiliation{Molecular Foundry, Lawrence Berkeley National Laboratory, Berkeley, CA 94720, USA}

\author{W. Hu}
\author{C. Mazzoli}
\author{S. B. Wilkins}
\affiliation{National Synchrotron Light Source II, Brookhaven National Laboratory, Upton, New York 11973, USA}

\author{I. K. Robinson}
\affiliation{Condensed Matter Physics and Materials Science Department, Brookhaven National Laboratory, Upton, New York 11973, USA}
\affiliation{London Centre for Nanotechnology, University College, Gower St., London, WC1E 6BT, UK}

\author{L. E. De Long}
\affiliation{Department of Physics, University of Kentucky, Lexington, KY 40506, USA}

\author{S. Roy}
\email{sroy@lbl.gov}
\affiliation{Advanced Light Source, Lawrence Berkeley National Laboratory, Berkeley, CA 94720, USA}

\author{J.T. Hastings}
\email{todd.hastings@uky.edu}
\affiliation{Department of Electrical and Computer Engineering, University of Kentucky, Lexington, KY 40506, USA}

\newcommand{\microns}{$\mathrm{\mu}$m}
\newcommand{\angstrom}{\mbox{\normalfont\AA}}
\date{\today}

\begin{abstract}

Collective dynamics often play an important role in determining the stability of ground states for both naturally occurring materials and metamaterials. We studied the temperature dependent dynamics of antiferromagnetically ordered superdomains in a square artificial spin lattice  using soft x-ray photon correlation spectroscopy.  We observed an exponential slowing down of superdomain wall motion below the AF onset temperature, similar to the behavior of typical bulk antiferromagnets. Using a continuous time random walk model we show that these superdomain walls undergo  low-temperature ballistic and high-temperature diffusive motions.

\end{abstract}



\maketitle
Naturally occurring systems with dipole magnetic interactions exhibit exotic emergent phases, such as quantum spin liquids \cite{Sibille, Balents2010}, and novel magnetic excitations \cite{morris2009dirac}.  Fluctuations about equilibrium in such systems are inevitable and remain incompletely understood. Moreover, low phase transition temperatures and lack of control in engineering the energy landscapes of atomic systems pose significant challenges to understanding the fundamental physics underlying spin ice behavior. Artificially fabricated lattices mitigate these problems and have attracted increasing attention as appropriate model systems for elucidation of frustration, phase transitions and associated dynamics \cite{Mengotti2011, Morgan2011, Wang2016, Ostman}. 

Artificially fabricated lattices commonly consist of dipole-coupled, elongated, nanoscale segments of ferromagnetic thin films (``block-spins'') placed on a two-dimensional periodic lattice. The shape anisotropy of the block-spins constrains their magnetization to lie along their long axis, which creates a classical analog of Ising spins. We refer to such systems as `artificial spin lattices' (ASL), which includes the intensively studied artificial spin ices \cite{lammert2012gibbsianizing}. In particular, a 2D square ASL exhibits an antiferromagnetic ground state \cite{Moller, Nisoli2007,Morgan2011,kapaklis2012melting,Farhan2013,Kapaklis2014, Andersson2016}, whose simple structure serves as an ideal model system for studies of equilibrium dynamics in dipolar-coupled systems. 

Previous investigations of thermally-active, square ASL indicate that a magnetic phase transition from an ordered antiferromagnetic (AF) ground state to a disordered paramagnetic (PM) state takes place at a temperature, $T_N$.  Large AF domains form well below $T_N$ \cite{kapaklis2012melting, Kapaklis2014}. Such mesoscopic domains are referred to as superdomains to distinguish them from microscopic domains in the magnetic thin-film \cite{Super}.  When the temperature approaches $T_N$, the system forms contiguous regions of rapidly fluctuating block-spins coexisting with AF superdomains (See reference \cite{Kapaklis2014} and Fig.~S6(a) in the Supplemental Information).  

Static AF superdomains in square ASL have been imaged using magnetic force microscopy (MFM) \cite{Morgan2011,ZhangSheng2013Comc}.  In one case, frozen thermal excitations above the AF ground state were observed within these superdomains \cite{Morgan2011}.  Lorentz transmission electron microscopy has also been used to image similar static superdomains in square ASL with topological defects \cite{Drisko2017}.  On the other hand, dynamics in the square ASL have been imaged with photoemission electron microscopy (PEEM) to study fluctuations of individual block-spins \cite{Kapaklis2014} and relaxation from ferromagnetic states \cite{Farhan2013}. However, these studies did not capture the collective fluctuations of block-spins at superdomain boundaries.  Moreover, these studies were limited to the PEEM time resolution of a few seconds \cite{Kapaklis2014, Farhan2013, gilbert2016emergent,faststrobePEEM}.

Here we report the direct observation of spontaneous AF superdomain wall nucleation, annihilation, and fluctuations in a 2D square ASL.  We have used resonant coherent x-ray diffraction over a wide range of temperatures near the AF-to-PM phase transition.  Coherent x-rays can directly probe order parameters and collective dynamics.  The diffraction pattern of coherent x-rays from magnetic domains includes a complex interference (speckle) pattern that is unique to the real-space superdomain textures. By tracking time-dependent speckle motion, we studied superdomain dynamics in square ASL with 100-millisecond time resolution. We applied a random-walk model that revealed two distinct regimes of superdomain wall motion as the sample goes through the AF phase transition: a low-temperature ballistic and a high-temperature diffusive type.  These studies show that characterizing superdomain wall behavior is critical to understanding the dynamics of the square ASL.  Such an understanding may prove crucial for implementing computing and data storage strategies based upon artificial spin systems \cite{jensen2018computation,arava2019engineering,arava2018computational,hehn2018bio}.


\vspace{2mm}
\begin{figure}
    \includegraphics[width=\linewidth]{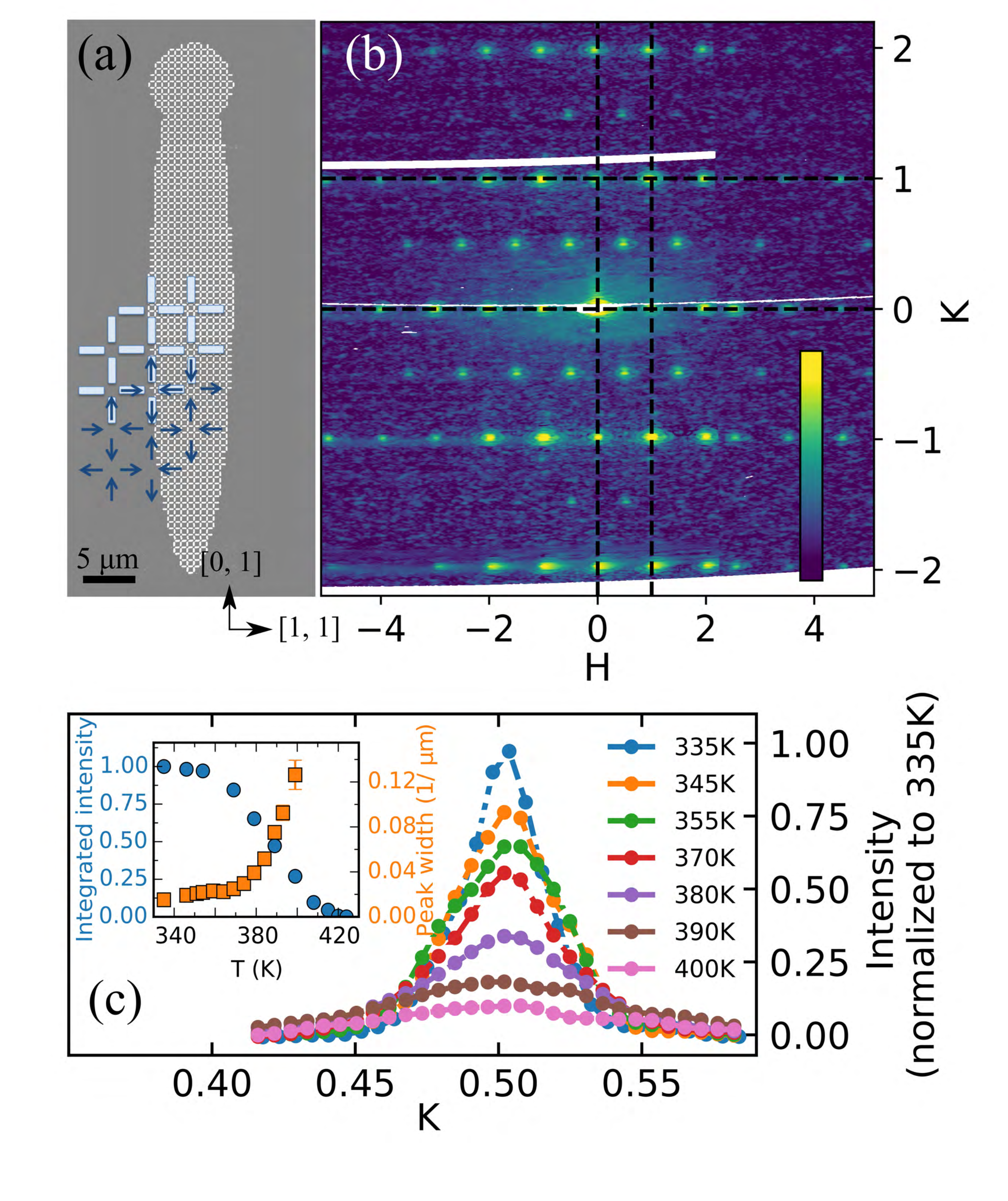}
    \caption{(a) SEM image of a square ASL sample with a schematic of the block-spin lattice and AF-ordered block-spins. (b) ASL diffraction pattern ($T = 335~K$) in reciprocal lattice coordinates $\left(H, K\right) =  \left(Q_x/\left(2\pi/a\right), Q_y/\left(2\pi/b\right)\right)$, where $a$ = $b$ = lattice constant (600~nm). Detector intensity is plotted on a log scale in arbitrary units (note color scale). A half-integer AF Bragg peak is clearly visible at the center of the box bounded by black dotted lines.  (c) Temperature dependence of detector intensity along a cut through time-averaged AF Bragg peaks. Inset shows the integrated intensity and peak width obtained from a Lorentzian fit.
    }
    \label{set_up_and_I}
\end{figure}
A square permalloy (Ni$_{0.8}$Fe$_{0.2}$) ASL was fabricated on a silicon nitride membrane using electron-beam lithography (Fig.~\ref{set_up_and_I}~(a)). The block-spin dimensions were 470~nm long, 170~nm wide, and 3~nm thick with a lattice constant $a = $ 600~nm. Figure 1(a) also illustrates the AF ground state configuration of the ASL.  Coherent x-ray diffraction measurements were performed at Beamline 23-ID-1 at the National Synchrotron Light Source II. 
The sample was positioned at a glancing angle of $\theta = 10^{\circ}$ with respect to the $\sigma$-polarized beam propagation direction to enhance the in-plane x-ray magnetic cross-section with the [0, 1] axis in the scattering plane. The detector was centered on the specularly (zero order) reflected beam. The sample's elongated shape ($\sim$~8~$\mu$m~x~50~$\mu$m), tailored for the $10^{\circ}$ glancing angle, maximizes the scattering volume while satisfying the Nyquist sampling condition.  Essentially perfect transverse coherence and a longitudinal coherence length of $\sim$~2~$\mu$m are realized at the sample. The incident x-ray beam always overfills the sample area to minimize artifacts from beam drift or from shifts in sample position due to temperature changes. Diffracted photons were collected using a fast CCD detector (with a readout rate of 10 Hz and 30~$\mu m$ x 30~$\mu m$ pixel size) placed 340~mm from the sample \cite{doering2011development}.

Figure~\ref{set_up_and_I}~(b) shows a typical diffraction pattern from a square ASL in its AF ground state. Rows of intense structural Bragg peaks and  weaker AF Bragg peaks are visible at integer $\mathbi{q} = (H, K)$ and half-integer $(H/2, K/2)$ wavevectors, respectively, surrounding the central $(0, 0)$ specular reflection. The AF Bragg peaks could only be detected by using resonant enhancement from the magnetic Fe (Ni) edge at 707 (853)~eV \cite{S1}.  These peaks provide a direct measure of the strength and character of the AF order. Figure~\ref{set_up_and_I}~(c) shows cuts through time-averaged AF Bragg peaks at various temperatures spanning the magnetic transition. The inset plots the temperature dependence of the integrated AF Bragg peak intensity which corresponds to the area fraction of AF domains.  The width of the AF Bragg peak measures the AF correlation length; thus, we observe the AF superdomains shrink as temperature approaches the magnetic transition at $T_N \sim$ 425~K. Past theoretical studies \cite{Levis2013, xie2015magnetic} on square ASL predict a continuous phase transition; here, our observed transition is apparently broadened by a finite size effect \cite{finite_size, S2}.

\begin{figure*}
    \includegraphics[width=\textwidth]{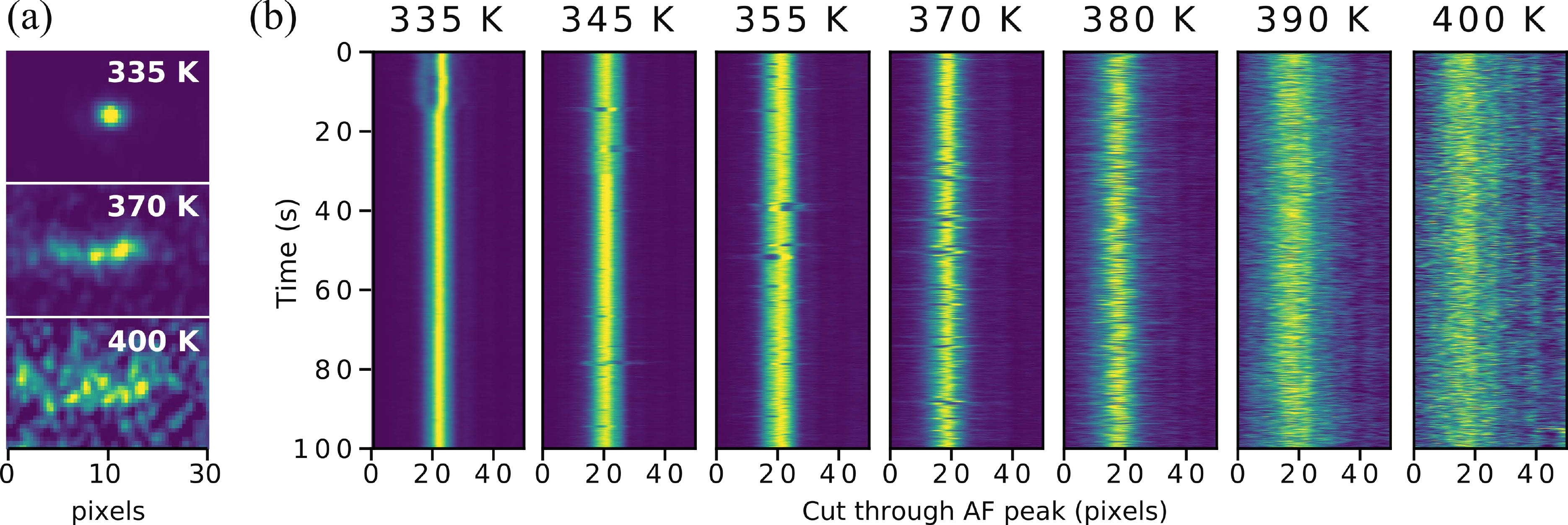}
    \caption{(a) Speckles observed for a single AF Bragg peak at three different temperatures. At 335 K we observe a single AF superdomain, and growth of speckle number with increasing temperature. (b) Waterfall plots showing the time evolution of speckle positions for various temperatures. Each horizontal line represents a cut through an AF peak capturing the intensity vs. pixel position at some time $t$. One pixel is approximately 0.005 in $K$.  Intensities are normalized to the maximum intensity for each temperature as given in Fig.~\ref{set_up_and_I}~(c). Spontaneous domain wall fluctuations are observed at all temperatures, but decrease in number with reduced temperature.}
    \label{speckles_and_waterfall}
\end{figure*}

The AF Bragg peaks show speckle patterns that arise from coherent interference between different AF superdomains (Fig.~\ref{speckles_and_waterfall}~(a)). This pattern reflects the square of the Fourier transform of the AF texture, and offers unique insights into the spatial character and dynamics of the AF state. The size and shape of speckles depend on the x-ray energy, sample illuminated area, and the scattering geometry \cite{attwood2017x}.  However, the number of speckles and spatial distribution of intensities are an  indication of the number of AF superdomains and their dynamics \cite{Ian_counting}. After initially heating above $T_N$ and then cooling until $T << T_N$, only a single speckle was observed in the AF peaks, consistent with the presence of only a single AF superdomain across the entire sample \cite{Airy}.  As we increase the temperature, thermal excitations nucleate superdomain walls that split a single speckle to multiple speckles. We note that single block-spin flips or multiple block-spin excitations\cite{Morgan2011} cannot create these speckle patterns which necessarily require multiple AF superdomains with distinct, extended boundaries.

To visualize the time evolution of speckle positions, we show `waterfall-plots' in Fig.~\ref{speckles_and_waterfall}~(b), which consist of speckle intensity measured along a vertical cut through the AF peak as a function of time with a resolution of 0.1~s. The fluctuation rate and number of speckles increases as the phase transition temperature is approached from below. This is a direct indication of the sample transitioning from a stable, single superdomain state to a highly-fluctuating, multidomain state. Spontaneous nucleation and annihilation of superdomain walls is apparent in the data shown in Fig.~\ref{speckles_and_waterfall}~(b). For example, at 345~K the system is initially in a single superdomain state that generates a single speckle. Around $t = 15$~s, the speckle splits into two, indicating a creation of a superdomain wall. Subsequently at $t = 16$~s, the system evolves back to a single superdomain.  This collective behavior was not considered in prior PEEM studies that focused on single block-spin fluctuations \cite{Kapaklis2014} and XPCS studies that focused on weakly interacting block-spins \cite{Morley2017}. Such spontaneous behavior clearly arises from an equilibrium fluctuation instead of the magnetic relaxation processes previously observed by PEEM \cite{Farhan2013}.

\begin{figure}
    \includegraphics[width=0.45\textwidth]{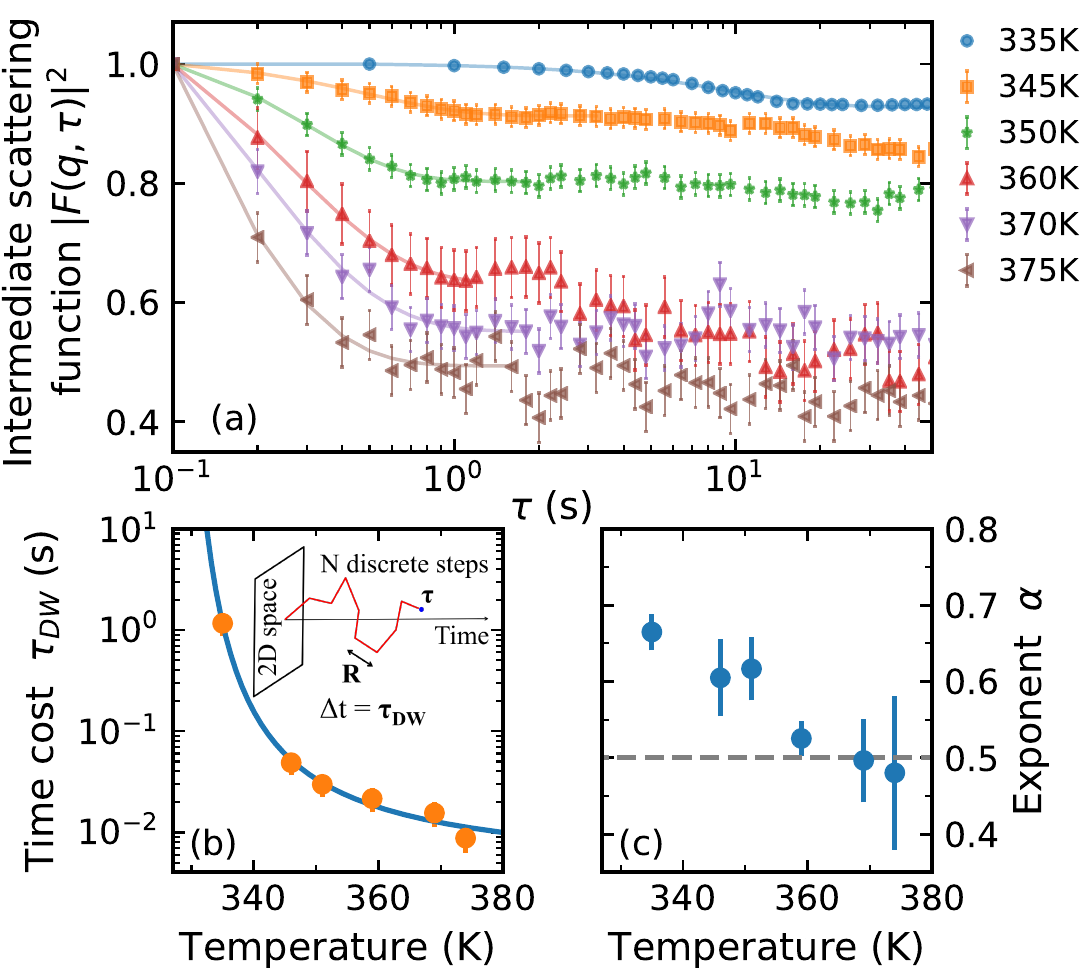}
    \caption{ (a) Intermediate scattering function $|F|^2$ calculated for speckle patterns at different temperatures. Solid lines are random walk fits to the initial decay. Temperature dependence of (b) time cost, $\tau_{\text{DW}}$, and (c) exponent, $\alpha$, in the model. (Inset in (b)) Random walk model schematic.}
    \label{g2_RWfit}
\end{figure}

The speckle time dependence is quantified using the one-time correlation function, $g_2(\mathbi{q}, \tau)$, given by
\begin{equation}
    g_2(\mathbi{q}, \tau) = \frac{\langle I(\mathbi{q},t)I(\mathbi{q}, t+\tau)\rangle}{\langle I(\mathbi{q}, t)\rangle ^2} = 1 + \beta \vert F(\mathbi{q},\tau)\vert ^2.
    \label{eq:g2}
\end{equation}
where $I(\mathbi{q},t)$ is the total intensity of a speckle image at wavevector $\mathbi{q}$ and at time $t$. The brackets $\langle\rangle$ indicate the time and ensemble average over all speckles with equivalent \mathbi{q} values. We can further write $ g_2(\mathbi{q}, \tau)$ in terms of the intermediate scattering function $\vert F(\mathbi{q},\tau)\vert ^2$ of the sample, and speckle contrast $\beta$ that depends only on the experimental setup \cite{beta,Chen2016}.  We calculate $\vert F(\tau)\vert ^2$ using detector areas corresponding to a single speckle.
Figure~\ref{g2_RWfit}~(a) shows that the decay time clearly decreases with increasing temperature.  Above 385~K , $\vert F(\tau)\vert ^2$ flattens \cite{S3}, as the fluctuations become faster than the CCD acquisition rate. We did not observe a clear $\mathbi{q}$-dependence of the speckle correlation up to 375K, as the speckle intensity drops sharply with increasing $\mathbi{q}$. In principle, $\vert F(\tau)\vert ^2$ drops from $1$ to $0$ upon complete decorrelation of a speckle pattern. In our case, $\vert F(\tau)\vert ^2$ drops to a temperature-dependent, finite offset that depends on the static fraction of AF superdomains. 

The dramatic temperature dependence in the curvature of $\vert F(\tau)\vert ^2$ indicates a change in the nature of superdomain dynamics. To understand this behavior, we developed a model that maps magnetic superdomains onto particles positioned at the center of mass of the superdomain boundaries \cite{S4}.  In this approach, we are not sensitive to fluctuations of individual block-spins, but our model adequately describes $\vert F(\tau)\vert ^2$ because our signal is dominated by speckles in low-$q$ regions.  Movements of dilute particles in media are often modeled with continuous time random walk (CTRW) behavior \cite{montroll1965random, duri2006length, Caronna} where the de-correlation of speckles at $(q, \tau)$ is the expected value of the degree of correlation $h$ weighted by its probability density function (PDF) $P_{\tau_{\text{DW}}}$ such that 
\begin{equation}
    F\left(q,\tau\right) = \sum^\infty_{N = 0} P_{\tau_{\text{DW}}}\left(\tau, N\right) h\left(q, N\right).
    \label{eq:random_walk}
\end{equation}
We take $P_{\tau_{\text{DW}}}\left(\tau, N\right)$ to be a Poisson distribution $\left(\tau/\tau_{DW}\right)^N e^{-\tau/\tau_{DW}} /N!$ describing the probability density of the number of steps $N$ that a particle traveled in time $\tau$, with variable time cost $\tau_{\text{DW}}$ between each step (Fig.~\ref{g2_RWfit}~(b)~inset) \cite{Metzler2000, bouchaud1990anomalous}. When averaged over all domains and traveling directions, one can write $h\left(q, N\right)\sim\text{exp}(-\left(q R N^{\alpha}\right)^2)$, assuming a constant displacement $R$ of superdomain boundaries during each step \cite{duri2006length}. 

Here we used $R \sim$ 0.8$a$, the center of the PDF of domain boundary displacements for a single jump, where $a$ is the lattice parameter of square ASL \cite{S5}. The exponent $\alpha$ describes the nature of the particle motion and ranges from 0 to 1. Two regimes, $\alpha < 1/2$ and $\alpha > 1/2$, correspond to sub-diffusion and hyper-diffusion respectively.  There are two special cases:  $\alpha = 1$ describes unidirectional motion over the decorrelation time of the system, commonly referred to as `ballistic' motion, and $\alpha = 1/2$ describes Brownian motion. In Fig.~\ref{g2_RWfit}~(b) and (c), we plot the temperature dependence of $\tau_{\text{DW}}$ and $\alpha$ obtained by individually fitting $F\left(q,\tau\right)$ with Eq.~\ref{eq:random_walk} for fixed $q$. Here, the initial decay in $F\left(q,\tau\right)$, fit to delay times of $\approx 2$s, provides insight into the block-spin collective dynamics. Our analysis does not eliminate the possible existence of faster ($\tau<0.1$~sec) or slower dynamics occuring beyond the initial decay.

In Fig.~\ref{g2_RWfit}~(c), the exponent $\alpha$ starts off close to $0.65$ at 335~K and drops to $0.5$ as temperature approaches to $T_N$, suggesting that the nature of superdomain motion changes from ballistic to diffusive. This can be explained considering two types of domain boundaries: superdomain walls separating two AF superdomains or phase boundaries separating AF superdomains and paramagnetic regions. Consider, for example, an initially AF-ordered ground state that encompasses the whole sample. When a superdomain wall spontaneously nucleates, the system tries to minimize the energy by pushing the superdomain wall out of the sample. The superdomain wall travels until it is scattered by another wall, is pinned by a defect, or reaches the sample's edge. Therefore, at low temperatures, superdomain walls appear to behave ballistically. On the other hand, at high temperatures, the sample is broken into small AF superdomains separated by paramagnetic regions. In this regime, each AF superdomain can move independently with no additional energy cost, and therefore the phase boundaries exhibit diffusive motion. This interpretation is also consistent with a continuous phase transition in which boundary effects lead to phase separation \cite{cugliandolo2017artificial}.

The characteristic time cost for domain wall motion $\tau_{\text{DW}}$ increases at low temperatures and diverges as superdomain walls freeze at a singularity, as shown in Fig.~\ref{g2_RWfit}~(b). This type of behavior is often described using the Vogel-Fulcher-Tammann (VFT) law in systems dominated by domains \cite{Tarjus2005, Mallamace}:
\begin{equation}
    \tau_{\text{DW}} = \tau_o~\text{exp}\left(DT_o/\left(T-To\right)\right).
    \label{eq:VFT}
\end{equation}
where $T_o$ is the freezing-in temperature and $D$ is the fragility of the system.  The smaller the ``fragility,'' the more the system deviates from an Arrhenius-type behavior.  

The decay time, $\tau_{\text{DW}}$ in Eq. \ref{eq:VFT}, is well fitted using $\tau_o = 0.003(2)~s$, $D = 0.17(6)$ and $T_o = 326(2)~K$ (solid line in Fig.~\ref{g2_RWfit}~(b)), indicating that the superdomain wall movement exponentially slows as $T$ approaches $T_o$ \cite{blocking}. The value of $D$ obtained is surprisingly similar to that of magnetic domains of a spiral antiferromagnet ($D = 0.14$ \cite{chen2013jamming}).  Our $\tau_o$, the characteristic fluctuation time as $T \rightarrow \infty$, is large compared to values observed for nanoparticles ($\approx 10^{-10}~$s) \cite{Brown1979a}. This is consistent with the nature of superdomain boundary fluctuations that necessarily require multiple block-spin flips.  If we consider individual block-spin flips in the limit that $T \gg T_o$, we find $\tau_{\text{DW}}$ is well modelled by fluctuations involving approximately 4 block-spins. (See Supplementary Material section S6.) This result is consistent with AF domains fluctuating by one lattice unit cell when surrounded by PM regions.

\begin{figure}
    \includegraphics[width=0.45\textwidth]{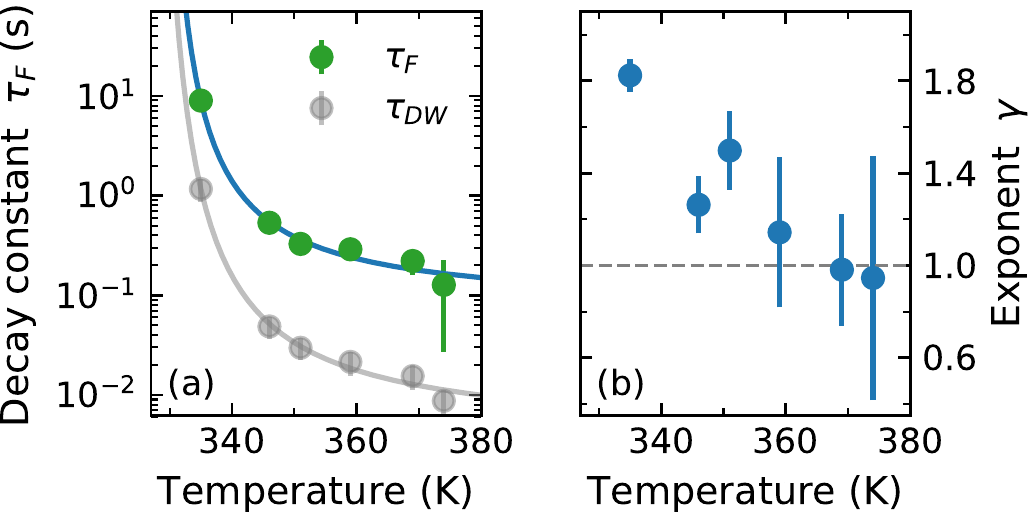}
    \caption{(a) Temperature dependence of decay time $\tau_{\text{F}}$, compared with $\tau_{\text{DW}}$. (b) Compressed exponent $\gamma$.}
    \label{stretched_exp}
\end{figure}
Finally, we compare our random walk model to a stretched exponential function: $F\left(\tau\right) = a~\text{exp} \left(-\left(\tau/\tau_{\text{F}}\right)^\gamma \right) + (1-a)$
that is commonly employed to understand XPCS data for collective phenomena in glasses and jammed systems \cite{chen2013jamming, Tarjus2005, Mallamace}. $\tau_{\text{F}}$ and $\gamma$ are the decay constant and stretched exponent, respectively, while $(1-a)$ accounts for the finite, temperature dependent offset explained earlier. Fig.~\ref{stretched_exp} compares the temperature dependence of $\tau_{\text{F}}$ obtained from the stretched exponential model to $\tau_{\text{DW}}$ obtained from the random walk model.  A VFT fit of the form in Eq.~\ref{eq:VFT} (solid lines) found that both models yield $D$ and $T_o$ values within the range of expected error.  The ratio of $\tau_o$ from the stretched exponential fit to $\tau_o$ from the CTRW model is $\sim 20$, comparable to the total number of lattice units across the sample. This suggests that $\tau_{\text{F}}$ is related to the travel time of a superdomain boundary (taking approximately 20 $\cdot$ $\tau_{\text{DW}}$ to move out of the sample). In addition, the exponent $\gamma$ decreases from 1.8 at $T = 335$~K to $\approx 1$ as the sample temperature approaches $T_N$ (Fig.~\ref{stretched_exp}~(b)). Our random walk model therefore gives a natural explanation for $\gamma$ where a compressed ($\gamma > 1$ ) and a simple ($\gamma = 1$) exponential indicate collective and diffusive motion of superdomain boundaries, respectively.

In summary, resonant coherent x-ray scattering provides unique insights for understanding the equilibrium behavior of a square ASL near its AF-to-PM phase transition temperature $T_N$.  As temperature decreases below $T_N$, AF superdomain sizes increase and magnetic fluctuations slow.  Applying both CTRW and stretched exponential models to the time correlation of the AF speckle pattern revealed a dynamical crossover temperature below $T_N$ near which superdomain wall motion changes from diffusive to ballistic.  Below this crossover temperature, the superdomain walls exponentially slow down with decreasing temperature and freeze in at $T_o$ as determined by the VFT model.   

These results show that superdomain-wall nucleation, annihilation, and motion are important for governing the complex equilibrium fluctuations of square artificial spin lattices. The methods described here can be readily applied to studies of the effects of disorder and defects in various artificial lattices \cite{gartside2018realization, savary2017disorder, sendetskyi2016magnetic,shen2012dynamics, glavic2018spin}. Similar collective motion of spins likely exists in other phase separated materials and could be explored using coherent x-rays \cite{chen2013jamming,tao2016direct}. Moreover, our findings concerning equilibrium fluctuations may prove important when engineering ASL for infromation technology or other applications \cite{neumann2010quantum,jensen2018computation,arava2019engineering,arava2018computational,hehn2018bio}. 

\begin{acknowledgments}
This material is based upon work is supported by the U.S. Department of Energy, Office of Science, Office of Basic Energy Sciences under Award Number DE-SC0016519.
\end{acknowledgments}

\bibliography{refs_ASI}

\begin{thebibliography}{56}
\bibitem [{Air()}]{Airy}%
  \BibitemOpen
  \href@noop {} {}\bibinfo {note} {A single superdomain is equivalent to a
  `negative' pinhole, which will create an Airy ring-like diffraction pattern.
  The peak we see is, in fact, the Airy ring center, whose width doubles a
  normal speckle size (or the Airy ring fringe width).}\BibitemShut {Stop}%
\bibitem [{blo()}]{blocking}%
  \BibitemOpen
  \href@noop {} {}\bibinfo {note} {We note here that $T_o$ should not be
  confused with the blocking temperature of isolated block-spins, which is at
  much lower temperature.}\BibitemShut {Stop}%
\end{thebibliography}
\end{document}


\title{Supplemental Material for: Spontaneous Magnetic Superdomain Wall Fluctuations in an Artificial Antiferromagnet}

\renewcommand{\thefigure}{S\arabic{figure}}
\renewcommand{\thesection}{S\arabic{section}}
\renewcommand{\theequation}{S\arabic{equation}}

\date{\today}

\maketitle



\section{Energy dependence of AF Bragg peak intensity\label{sec:Escan}}
\begin{figure}[h!]
    \includegraphics[width=0.4\textwidth]{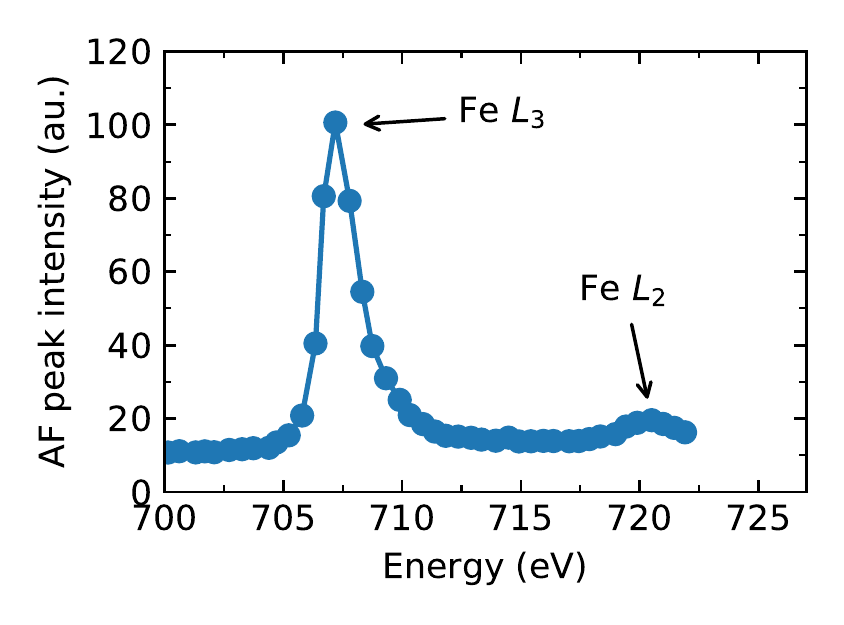}
    \caption{Energy dependence of the AF Bragg peak intensity. A strong resonance at $L_3$ edge and a weak resonance at $L_2$ edge are clearly visible.}
    \label{Sup_Escan}
\end{figure}

Fig.~\ref{Sup_Escan} shows the energy dependence of the AF Bragg peak intensity. The AF peak appears only at the Fe~$L_3$ absorption edge, providing a direct measure of the strength of the AF order.

\section{Temperature history dependence of AF Bragg peak intensity}
Theoretical calculations predict a second-order magnetic transition for square ASL \cite{Levis2013}. However, our experimental results and those of others consistently indicate a broadened transition with a temperature hysteresis \cite{Andersson2016}, which can be attributed finite nature of artificial samples. 

\begin{figure}[h!]
    \includegraphics[width=0.4\textwidth]{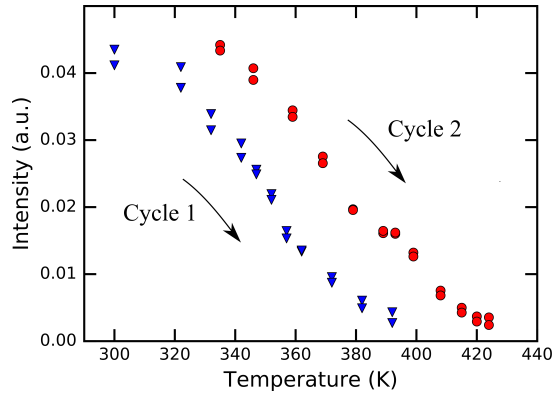}
    \caption{Temperature history dependence of AF Bragg peak intensity. Two data sets (in blue and red) are taken at separate temperature cycles on the same sample, both taken at increasing temperature.}
    \label{Sup_T_hyst}
\end{figure}

Fig.~\ref{Sup_T_hyst} shows the temperature dependence of the AF Bragg peak intensity taken for two separate times as the sample was warmed at different rate. The magnetic transition temperature is affected by the temperature history of the sample, consistent with earlier studies \cite{Andersson2016}.

%

\section{One-time correlation, \lowercase{$g_2$} for high temperature data sets}
\begin{figure}[h!]
    \includegraphics[width=0.45\textwidth]{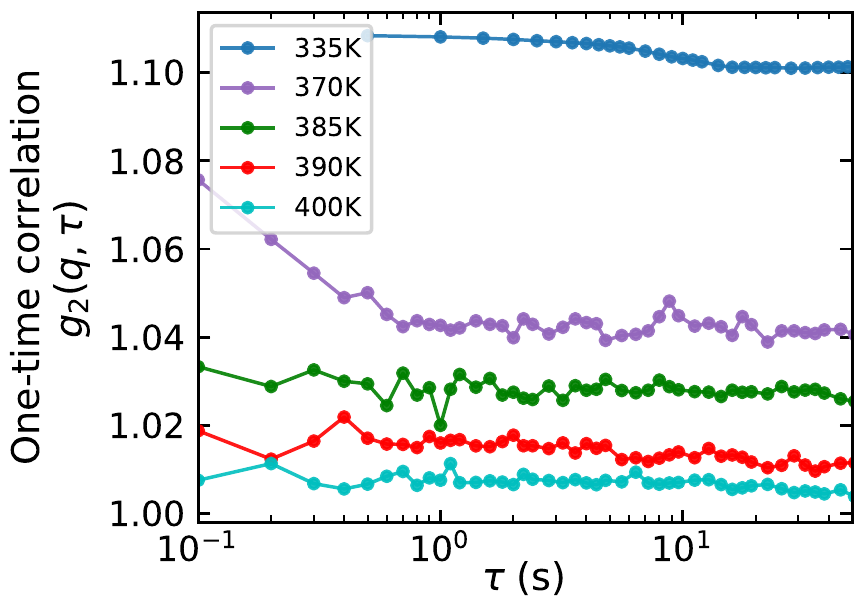}
    \caption{One-time correlation, g$_2$, calculated from speckles in AF Bragg peaks. High temperature data sets produce flat line due to time resolution of experiment, $i.e.$, speckle fluctuation is faster than fast-CCD camera count time.}
    \label{Sup_g2}
\end{figure}

One time correlation, $g_2$, becomes a constant when fluctuation is faster than the experiment's time resolution. In our case this corresponds to the data acquisition time of 0.1~s. In the main text, we showed $g_2$ curves with finite decays up to 375~K. In Fig.~\ref{Sup_g2}, we show $g_2$ calculated for higher temperatures (385, 390, 400~K) compared with two (slow and fast) decays included in Fig.3~(a). Fig.~\ref{Sup_g2} indicates that for all temperatures above 384K, super-domain wall fluctuation is faster than the data acquisition time. 

\section{Continuous time random walk model}
The continuous time random walk (CTRW) model is widely used to understand anomalous diffusion in various materials including electron transport in junctions, nanoparticles in liquid glass formers, and aging in colloidal gel \cite{nelson1999continuous,Caronna, guo2009nanoparticle,cipelletti2000universal}. It is a generalization of the random walk model where the wait time during each step is randomly sampled from a distribution. In the main text, we applied the CTRW model to understand the movement of superdomain boundaries. In this section, we map the lowest mode of the AF superdomain fluctuation into the motion of a diffuse particle positioned at the center of mass of the AF superdomain boundary, allowing us to use CTRW to understand its dynamics. In this formalism, the wait time during each step of the random walk is directly related to the energetic distribution of the dipole-interacting background.
\subsection{Fourier expansion of AF domain boundary}
\begin{figure}[h!]
    \includegraphics[width=0.5\textwidth]{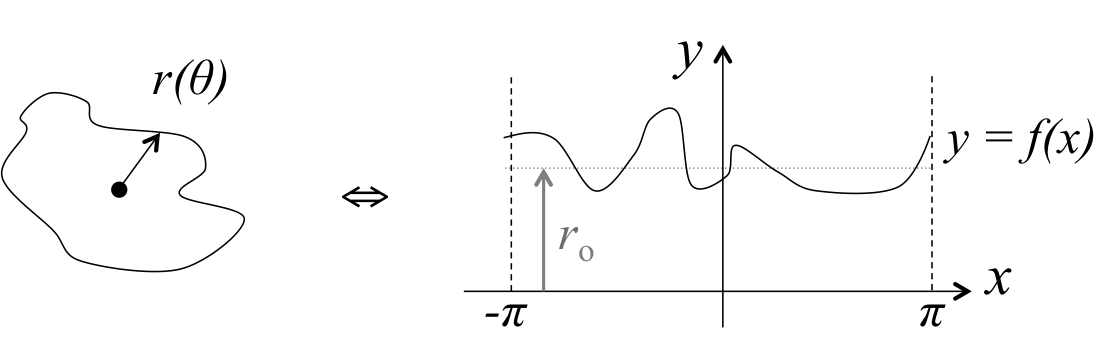}
    \caption{Transformation of a boundary to a string separated $r_o$ from x-axis.}
    \label{Sup_rw1}
\end{figure}
Consider an AF super-domain as in Fig.~\ref{Sup_rw1}. We consider the domain fluctuation in momentum space and therefore its Fourier expansion terms. Domain fluctuation is defined by domain boundary movement. Fourier expansion of the boundary $\mathbi{r}\left(\theta\right)$ is equivalent to Fourier expansion of a string at $y = f(x)$ with periodic boundary conditions of periodicity $2\pi$ as follows:
\begin{equation}
    \mathbi{r}\left(\theta\right) = f\left(x\right)\cos{x}|_{x=\theta}~\hat{\mathbi{i}} + f\left(x\right)\sin{x}|_{x=\theta}~\hat{\mathbi{j}}.
    \label{eq:cooperativity1}
\end{equation}
Generically, the Fourier expansion of a function $f(x)$ is given by
\begin{equation}
    f\left(x\right) = \frac{1}{2}a_o + \sum\limits_{n=1}^{\infty}a_n \cos{nx} + \sum\limits_{n=1}^{\infty} b_n \sin{nx}.
    \label{eq:cooperativity2}
\end{equation}
In our case, the lowest mode
\begin{equation}
    \frac{1}{2}\cdot a_o = \frac{1}{2}\cdot \frac{1}{\pi}\int_{-\pi}^{\pi}f\left(x\right)dx
    \label{eq:cooperativity3}
\end{equation}
represents a circle with radius $r_o$, at the center of mass (COM) of $\mathbi{r}\left(\theta\right)$.

\subsection{Distribution of superdomains}
\begin{figure}[h!]
    \includegraphics[width=0.25\textwidth]{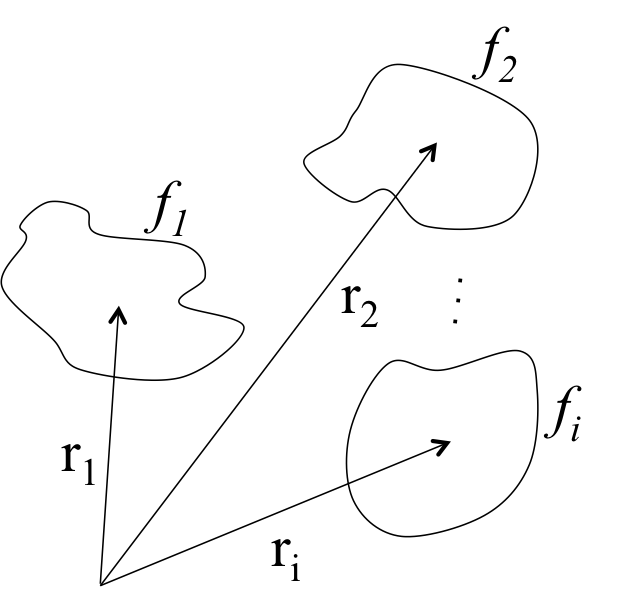}
    \caption{Density distribution of AF super-domains. $r_i$ represents position of domains at COM of domain boundaries. $f_i$ represents a function describing the shape of domain.}
    \label{Sup_rw2}
\end{figure}
A single domain at position $r_i$ with shape $f_i$ can be written as a convolution of $\delta(r_i)\circledast f_i$, where $\delta(r_i)$ is a delta function positioned at $r_i$. We consider the Fourier expansion of the superdomain distribution $\rho$ as in \ref{Sup_rw2}. The Fourier transform of $\rho$ for small $\mathbi{q}$ is given as the following:
\begin{equation}
\begin{split}
\mathscr{F} \left[\rho\left(\{\delta(r_i)\circledast f_i\}\right)\right] 
&= \mathscr{F}  \left[\rho\left(\{\delta(r_i)\circledast (\frac{1}{2}a_o + \sum\limits_{n=1}^{\infty}a_n \cos{nx} + \sum\limits_{n=1}^{\infty} b_n \sin{nx})_i\}\right)\right]\\
& \sim \mathscr{F} \left[\rho\left(\{\delta(r_i)\circledast \frac{1}{2}a_{o,i}\}\right)\right] \\
& = \mathscr{F} \left[\rho\left(\{\delta(r_i)\circledast \TikCircle{}_{\left.r_{o}\right|_i}\}\right)\right]
\end{split}
\end{equation}
where $\{\}$ represents a set including all superdomains in the sample with individual elements $i$. The last step uses the result from section $\textbf{A}$, where the lowest mode of the Fourier expansion is a circle, $\TikCircle{}_{\left.r_{o}\right|_i}$, of radius $r_o$ for each domain $i$. This approximation is possible because we did not see any $q$-dependence in our data, and therefore higher order terms can be neglected. Normally, these terms cannot be ignored. We can approximate $\left.r_{o}\right|_i$ as average domain size $\xi$ and write
\begin{equation}
\begin{split}
\mathscr{F} \left[\rho\left(\{\delta(r_i)\circledast \TikCircle{}_{\left.r_{o}\right|_i}\}\right)\right]\\
& = \mathscr{F} \left[\rho\left(\{\delta(r_i)\circledast \TikCircle{}_\xi\}\right)\right]\\
& = \mathscr{F} \left[\rho\left(\{\delta(r_i)\}\right)\circledast \TikCircle{}_\xi\right]\\
& = \mathscr{F} \left[\rho\left(\{\delta(r_i)\}\right)\right] \cdot \mathscr{F} \left[\TikCircle{}_\xi\right].
\end{split}
\end{equation}
The first term in multiplication is Fourier transform of dilute particles positioned at the COM of the AF superdomain boundaries. The second term is the AF Bragg peak envelope. Therefore, speckle pattern at low-$\mathbi{q}$ region can be approximated to those from diffuse particle distributions. 

\section{Displacement $R$ of superdomain boundaries during each step used in CTRW model}
\begin{figure}[h!]
    \includegraphics[width=0.75\textwidth]{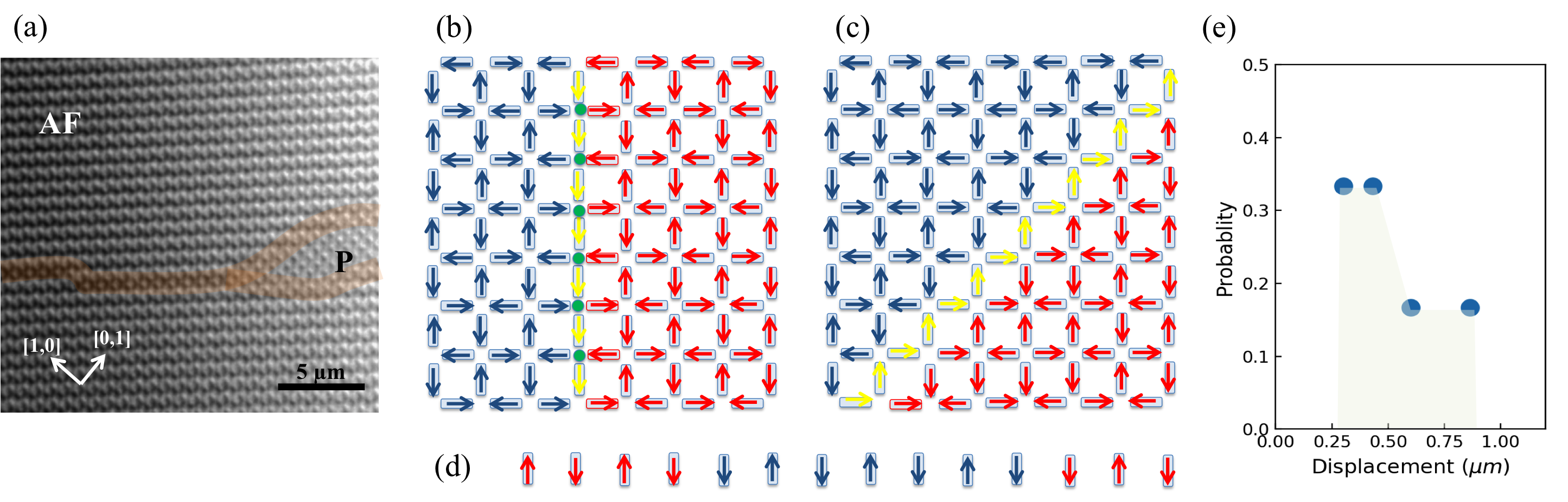}
    \caption{(a) Image of superdomain boundaries in a 100~$\mu m$~x~100~$\mu m$ square ASL sample obtained by x-ray photoemission electron microscopy. Two types of domain boundaries are highlighted with transparent orange lines: one separating two AF susperdomains (AF), and two separating AF superdomains and a fast-fluctuating region (P). Image was taken on our own sample fabricated with the same condition and lattice parameters except for the pattern size. Data were taken at beamline 11.0.1 of the Advanced Light Source with accumulation time of approximately 30 seconds. (b)(c) Two types of superdomain boundaries: vertical and diagonal. When all the block-spins in the boundary moves in one direction, this is equivalent to 1D spin chain in (d). (e) Probability distribution function for single step displacement of 1D superdomain wall}
    \label{Sup_R}
\end{figure}

We approximated the value of $R$ under the following assumptions.
Assumption 1: Superdomain walls like to form in a smooth string-like manner, in order to minimize the sample's global energy. This behavior is observed for superdomain walls with slow dynamics using x-ray photoemission electron microscopy \ref{Sup_R}~(a).
Assumption 2: During each random walk step, the superdomain walls prefer to move to the lower or equal energy state. Any local buckling of the superdomain wall creates an extra energy cost. A simultaneous shift of all block-spins in the superdomain wall produce the same energy for a large sample.
Assumption 3: Because of the elongated shape of the sample, superdomain walls are likely to travel along the long-axis of the sample since it costs less energy. Note that local fast fluctuations can create superdomain walls that do not follow these assumptions. However, these fluctuation exceed our time resolution or the $q$-range.

Using the assumptions above, we consider a single-step translation of domain walls in the CTRW model in both vertical and diagonal direction as in Fig.~\ref{Sup_R}~(b), (c). These two cases include the shortest and longest displacement of the superdomain walls. For simplicity, we consider one dimensional case with two boundaries as in (d). Then, the probability distribution for possible displacement is plotted in (d), with expectation value of $R = 0.8a$, where $a=600~nm$ is square ASL lattice parameter. We used $R = 0.8a$ to fit exponent $\alpha$ in Fig.3~(e), but using any $R$ in the distribution fits well inside the error bars. Errors mainly come from the $q$-range used to calculate $F\left(q,\tau\right)$. Finally, when our assumption fails, the value of $R$ can certainly change. However, a change in $R$ only results to a slight change in the the slope of $\alpha$, but does not affect our conclusion concerning  superdomain wall behavior. 

\section{Block-spin fluctuations as $T \rightarrow T_N$}
As the sample temperature approaches $T_N$ the AF domains are surrounded by fast-fluctuating PM regions.  This is also a regime where $T \gg To$ in the VFT model presented in Eq. 3.  If we assume block-spin flips are independent at these high temperatures then we can write the VFT equation as 
\begin{equation}
    \tau_{\text{DW}} = (\tau_0'~\text{exp}\left(-E_o/T\right))^{N_s},
    \label{eq:cooperativity4}
\end{equation}
where $\tau_0' \approx 10^{-10}~s$\cite{Brown1979a} is the fluctuation time for a single block-spin.  $N_s$ is the average number of block-spins flips involved in boundary motion, and $E_o$ is the energy cost of a single block-spin flip. In this formalism, $\tau_{\text{DW}}$ is well fitted by $N_s \sim 4$, the number of block-spins in the lattice unit cell.  Thus, as $T \rightarrow T_N$, the scale of fluctuations of the superdomains is consistent with the smallest possible change in their boundaries.

\bibliography{refs_ASI}